\documentclass[prl,aps,twocolumn,showpacs,superscriptaddress,reprint]{revtex4-1}
\usepackage{}
\usepackage{amsmath}
\usepackage{graphicx}
\usepackage{booktabs}
\usepackage{bm}
\usepackage{amssymb}
\usepackage{hyperref}

\begin{document}

\title{Coupling motion of colloidal particles in quasi-two-dimensional confinement}

\author{Jun Ma}
\email[J. Ma:  ]{junma@nwu.edu.cn, jun.ma35@outlook.com}
\affiliation{Department of Physics, Northwest University, 710069, Xian, China}
\author{Guangyin Jing}
\email[G. Jing:  ]{jing@nwu.edu.cn, guangyin.jing@gmail.com}
\affiliation{Department of Physics, Northwest University, 710069, Xian, China}
\affiliation{NanoBiophotonics center, National Key Laboratory and Incubation Base of Photoelectric Technology and Functional Materials, Xian, 710069, China}

\begin{abstract}
 Brownian motion of colloidal particles in the quasi-two-dimensional (qTD) confinement displays distinct kinetic characters from that in bulk. Here we experimentally report a dynamic evolution of Brownian particles in the qTD system. The dynamic system displays a quasi-equilibrium state of colloidal particles performing Brownian motion. In the quasi-equilibrium process, the qTD confinement results in the coupling of particle motions, which slowly dampens the motion and interaction of particles until the final equilibrium state reaches. The theory is developed to explain coupling motions of Brownian particles in the qTD confinement.
\end{abstract}

\maketitle
\begin{center}
\textbf{I. Introduction}
\end{center}

 Particles performing Brownian motion in quasi-two-dimensional (qTD) space ubiquitously exist in diverse systems in nature. These particles exhibit fundamentally different characterises from ones in bulk by the effects of the confinement and coupling interaction. Great potential applications recently stimulate extensively theoretical and experimental studies in various fields: self-propelled protein swimming in lipid bilayers of biological membranes~\citep{huang2012nano}, hydrodynamic diffusion of elastic capsules in bounded suspension~\citep{tan2012hydrodynamic}, hydrodynamic interactions between particles ~\citep{bonilla2012hydrodynamic,novikov2010hydrodynamic,uspal2012scattering} and particles flowing~\citep{C2SM25931A} in the suspension confined in the qTD channel, colloidal suspension transition to glass between two quasiparallel plates~\citep{nugent2007colloidal}, etc.

As colloidal suspension is confined between two quasiparallel glass surfaces, particle diffusion slows down for the reason that the regions for cooperative motion of particles in the confined space are qualitatively different than those in unconfined one~\citep{edmond2012influence}. The slower mobility mainly results from the confinement, with the dynamics of particles reducing dramatically as the separation distance (\emph{w}) of two surfaces decreasing~\citep{eral2009influence,carbajal1997brownian}. The reduction of one dimension causes the stronger hydrodynamic interaction between particles, the decreasing of the diffusion coefficient~\citep{carbajal1997brownian}. However, interestingly, this interaction declines with the distance (\emph{l}) of the pair of particles as $l^{-2}$, decaying much faster than that decreasing with $l^{-1}$ in the unconfined suspension~\citep{cui2004anomalous}.

The confinement leads to much larger interaction among particles than that in the bulk. On the other hand, the interaction and hydrodynamic coupling between Brownian particles and confined plates result in the reduction of mean squared displacement of particle~\citep{carbajal1997brownian}. The confinement also produces translational symmetry breaking of the suspension, thus the momentum of suspension is not conserved where the distance is larger than \emph{w}. The transverse momentum is a dominant contribution to the hydrodynamic interaction between the pair of particles in the bulk suspension, but is restrained in the confined one~\citep{diamant2009hydrodynamic}. The affluent distinct phenomena of particle suspension in confined geometries are however complicated and  far from well-understood.

In this work, we report an experimental finding on the existence of a quasi-equilibrium state of Brownian particles in a qTD system. Furthermore, we show the coupling motions for the particles, and develop the theory to describe its effect on Brownian motion.

\begin{center}
\textbf{II. Experiment and theoretical model}
\end{center}

Dilute colloidal suspension (volume fraction $1\%$) with the particle diameter of 1.39 micron is confined into two parallel quartz plates ($w=7.75 \mu m$). Heavy water is mixed into the suspension to eliminate the gravity effect. The motions of particles are observed and recorded by using optical microscopy under ambient condition. The trajectories of particles are analyzed from video by particle-tracking techniques~\citep{crocker1996methods}(see Appendix A for more experimental details).

Brownian motion of a free particle in the Ornstein-Uhlenbeck process is analyzed from the Langevin equation~\citep{uhlenbeck1930theory},
\begin{equation}\label{d1}
\frac{dv}{dt}+\beta v = A(t)
\end{equation}
 in which friction coefficient $\beta = 6\pi\eta_{s} r/m $ ($\eta_{s}$, viscosity of the suspension; $r$, particle radius; $m$, inertial mass of particle), $v$ is particle velocity, $t$ time, $A(t)=F/m$. $F$ is the stochastic force with zero mean, $\langle F(t)\rangle=0$; and its values are uncorrelated at different times ($t, t'$), $\langle F(t)F(t')\rangle = 12\pi\eta_{s} r k_{B}T\delta(t-t')$, $k_{B}$ is Boltzmann constant, $\emph{T}$ temperature, $\delta(t)$ Dirac delta function.

Particles performing Brownian motion in the suspension under a external force is described by Uhlenbeck and Ornstein as~\citep{uhlenbeck1930theory},
\begin{equation}\label{a1}
\frac{dv}{dt}+\beta v = A(t)+\frac{K(x)}{m}
\end{equation}
where $K(x)$ is the external force.

\begin{figure}[htbp]
\centering
\includegraphics[scale=0.6]{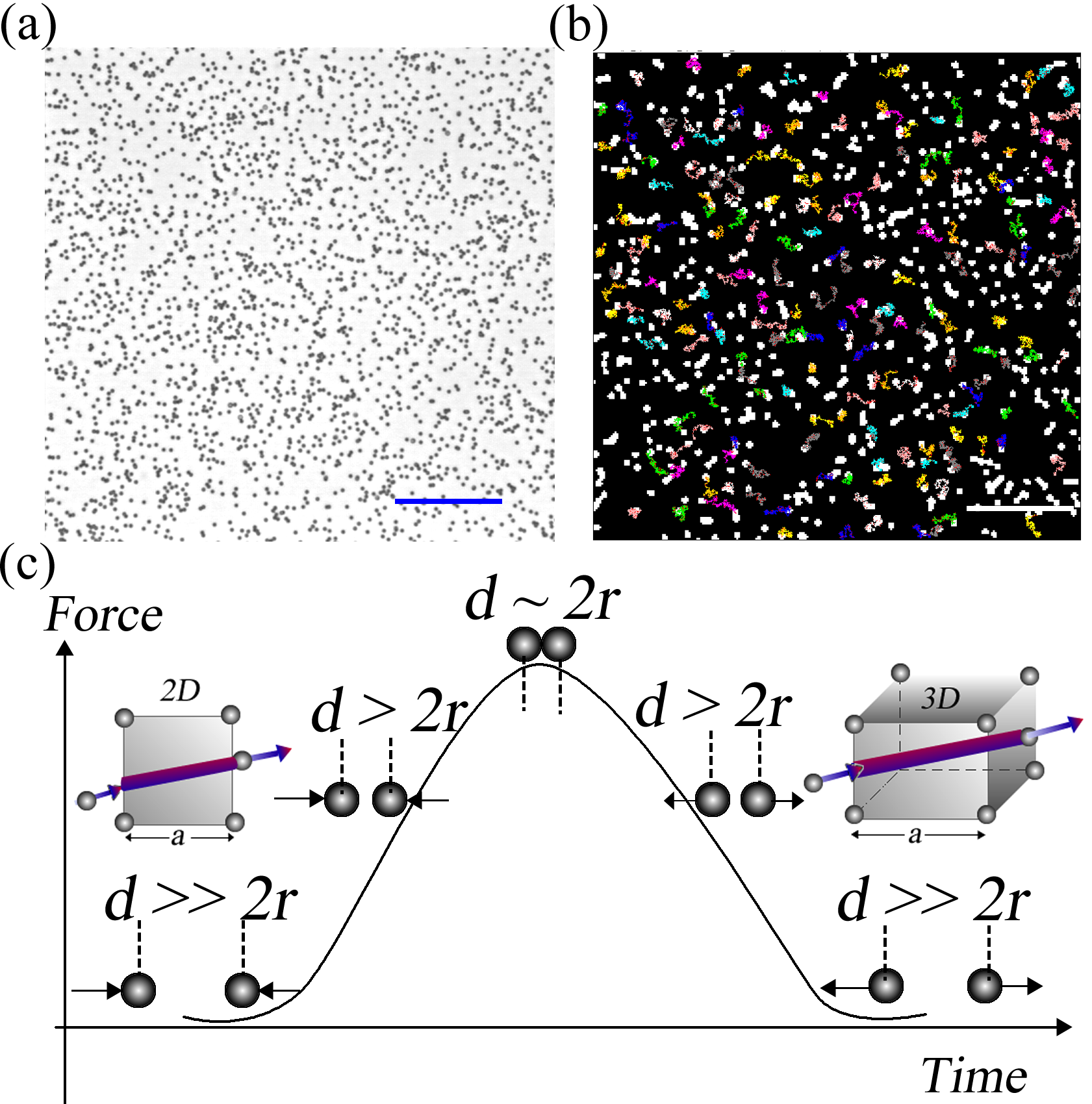}
\caption{Particles in the suspension performing Brownian motion and colliding to produce coupling interaction. (a) Optical micrograph of particle suspension confined in the qTD cell, the scale bar is $50 \mu m$.
(b) Typical trajectories of particle motions produced in 16.7 ms time interval. The picture is cropped from the upper left corner of Fig. 1(a) with the scale bar of $10 \mu m$.
(c) Sketch of the interaction of the particles depending on their separation. The distance \emph{d} between two particles can be much larger than, comparable and equal to the particle size, which corresponds to the week, moderate and strong interaction, respectively. A particle passing through a qTD area and 3D volume is illustrated in the left-hand and right-hand insets, respectively.}
\label{Fig1}
\end{figure}

Considering that colloidal particles (Fig. 1(a)) keep performing Brownian motion (Fig. 1(b)) in the present qTD cell, the coupling interaction among particles results from significantly increasing collision rate (number of collision/unit time) in the qTD confinement than that in the bulk (see Appendix B). Carbajal-Tinoco \emph{et al.} also indicated that in the confinement the hydrodynamic interaction among particles is much stronger than that in the bulk~\citep{carbajal1997brownian}. When a particle with a velocity $v$ is approaching the other one, the interaction force $F_{x}$ between them varies inversely as their distance $l$, \emph{i.e}. $F_{x}\varpropto 1/l$ ~\citep{batchelor1972hydrodynamic}. Batchelor and Green further indicated that as the two particles are nearly touching, a thin fluidic layer is produced between two particle surfaces. The strong interaction between two particles is acted by the stresses of large magnitude within a thin layer~\citep{batchelor1972hydrodynamic}.

After the collision, the two particles move apart with opposite direction (Fig. 1(c)). The fluidic layer between two particles thickens and then a liquid zone forms as two particles are pushed further apart. The force driving the particle apart gradually decreases to zero until it makes the particle reach the final velocity of $v'$ (Fig. 1(c)). The corresponding momentum change is, $\triangle P = mv'-mv$. The motion damping results from the energy loss of particles during the collision process. Here, we suppose that a virtual damping force acts in a collision process as, $\int f dt = mv'-mv = m(k-1)v$, where $v'=kv$, $k$ is velocity conversion ratio. Thus the damping force is, $f =m(k-1)(-dv/dt)$ (minus sign represents velocity dampening). In the process for approaching the equilibrium, since the velocity of particles decays with time, the energy loss in the collision process gradually decreases, \emph{i.e.} the damping force continually reduces as, $f_{d}=m(1-k)(dv/dt)e^{-\gamma t}$, $\gamma$ is the damping coefficient. Substitute $K(x)=f_{d}$ into Eq.\ref{a1}, we get the governing equation for the damping motion of Brownian particle in the qTD confinement,
\begin{equation}\label{a4}
[1+(k-1)e^{-\gamma t}]\frac{dv}{dt}+\beta v = A(t)
\end{equation}
The mean-square velocity $\langle v^{2}\rangle$ and the mean-square displacement $\langle(\triangle x)^{2}\rangle$ can be derived from Eq. \ref{a4} as (Appendix C),
\begin{equation}\label{a5}
\langle v^{2}\rangle=\frac{\alpha(t)k_{B}T}{m}(1-e^{-2\beta t})+\omega(t)v_{0}^{2}e^{-2\beta t}
\end{equation}
in which $\alpha(t)=[1+(k-1)e^{-\gamma t}]^{-2}, \omega(t)=[\frac{k}{1+(k-1)e^{-\gamma t}}]^{\frac{2\beta}{\gamma}}.$
\begin{equation}\label{a6}
\begin{array}{r@{~}l}
\langle(\triangle x)^{2}\rangle=&(\frac{k v_{0}}{\beta})^{2}[1-(\frac{k}{\gamma t+k})^{\frac{\beta}{\gamma}-1}]^{2}+ \frac{\sigma}{\beta^{2}[1+(k-1)e^{-\gamma t}]^{2}}\\
&[t+\frac{1}{2\beta}(-3+4e^{-\beta t}-e^{-2\beta t})]
\end{array}
\end{equation}
in which $\sigma=2\beta k_{B}T/m$.
For long time \emph{t}, Eqs.\ref{a5},\ref{a6} turn into the equations,
\begin{equation}\label{z1}
\langle v^{2}\rangle=\frac{\alpha(t)k_{B}T}{m}
\end{equation}

\begin{equation}\label{z2}
\langle(\triangle x)^{2}\rangle=\frac{\alpha(t)\sigma}{\beta^{2}}t
\end{equation}

For long time \emph{t}, comparing the mean-square velocity $\langle v_{ou}^{2}\rangle$ and the mean-square displacement$\langle(\triangle x)_{ou}^{2}\rangle$ derived from Eq.\ref{d1} in the Ornstein-Uhlenbeck process without external force~\citep{uhlenbeck1930theory}, $\langle v^{2}\rangle$ and $\langle(\triangle x)^{2}\rangle$ in Eqs. \ref{z1},\ref{z2} have an extra coefficient $\alpha(t)$. $\alpha(t)$ is considered as a damping factor, reflecting the damping effect resulting from coupling interaction of the particle motions. The damping factor $\alpha(t)$ is decreasing with the time \emph{t} increasing. It produces the dampening effect on the double aspects: $\langle v^{2}\rangle$ decreases with the time increasing; $\langle(\triangle x)^{2}\rangle$ increases with the time (\emph{t} in Eq.\ref{z2}) increasing, but the damping factor $\alpha(t)$ lessens the increasing rate. As $t\rightarrow \infty$, $\alpha(t)\rightarrow 1$, $\langle v_{\infty}^{2}\rangle = k_{B}T/m$ and $\langle(\triangle x)_{\infty}^{2}\rangle = \sigma t/\beta^{2}$, indicating that the damping effect finally vanishes and the coupling system of Brownian particles reaches an equilibrium.

\begin{center}
\textbf{III. Analysis of coupling motion of particles by theoretical model}
\end{center}

In the present qTD cell, particles are approximately treated as a monolayer, because usually the particle has not been observed to pass by the other particle above/beneath, instead, the collision always happens. These persistent collisions inevitably result in the damping of particle motion. The ratio of the distance between two quartz plates and particle size ($w/2r$) is 5.6, thus the viscosity of dilute suspension is the same as that of the bulk, in terms of the experimental study of Peyla \emph{et. al} ~\citep{peyla2011new}.

In the present system, during the slow process for approaching the equilibrium in qTD cell, lots of collisions have experienced for individual particle. In each collision process, the energy dissipation of the particle is very small amount. Therefore, the coupling interaction is weak dampening so that \emph{k} is close to unity and $\gamma$ is of small value. We get the root-mean-square (rms) velocity from Eq.\ref{z1} for the long dampening process,
\begin{equation}\label{x1}
v_{rms}= \sqrt{\langle v^{2}\rangle}=[1+(k-1)e^{-\gamma t}]^{-1}\sqrt{k_{B}T/m}
\end{equation}

As Einstein pointed out that the velocity and direction of particle motion change in extremely short time so that the instant velocity can hardly be measured~\citep{einstein1956investigations}. Here the motion of particles is measured with the absolute value of its displacement in a time interval $16.7 ms$ instead. The statistical result of huge quantities of trajectories displays a broad spectrum of the particle displacement $\triangle x$ (Fig. 2(a) inset). A distribution centering around 80 nm consists of huge quantities of the displacements and can been fitted with Gaussian function, which has maintaining during the coupling process (Fig. 2(b)).

Figure 2(a) indicates that the coupling interaction between particles leads to the damping of particle motion. The mean displacement $\langle\triangle x\rangle$ decreases from 139 nm at the beginning (1 h) to 131 nm at the equilibrium (100 h), with a $6\%$ reduction. Since the huge quantities of trajectory steps ($\sim10^{5}$) are considered here, this total reduction is a great amount, which is significant in the dynamic system. Average displacement $\langle\triangle x\rangle$ in a time interval $16.7 ms$, \emph{i.e.} average velocity $\langle\triangle v\rangle$, gradually decreases with the time, but the reduction rate decreases (Fig. 2(a)). It is consistent with the predication from Eq.\ref{x1} shown in Fig. 2(c). Therefore, the damping factor $\alpha(t)$ in Eq.\ref{z1} can describe particle motion slowly damping into the final equilibrium.
\begin{figure}[htbp]
\centering
\includegraphics[scale=0.5]{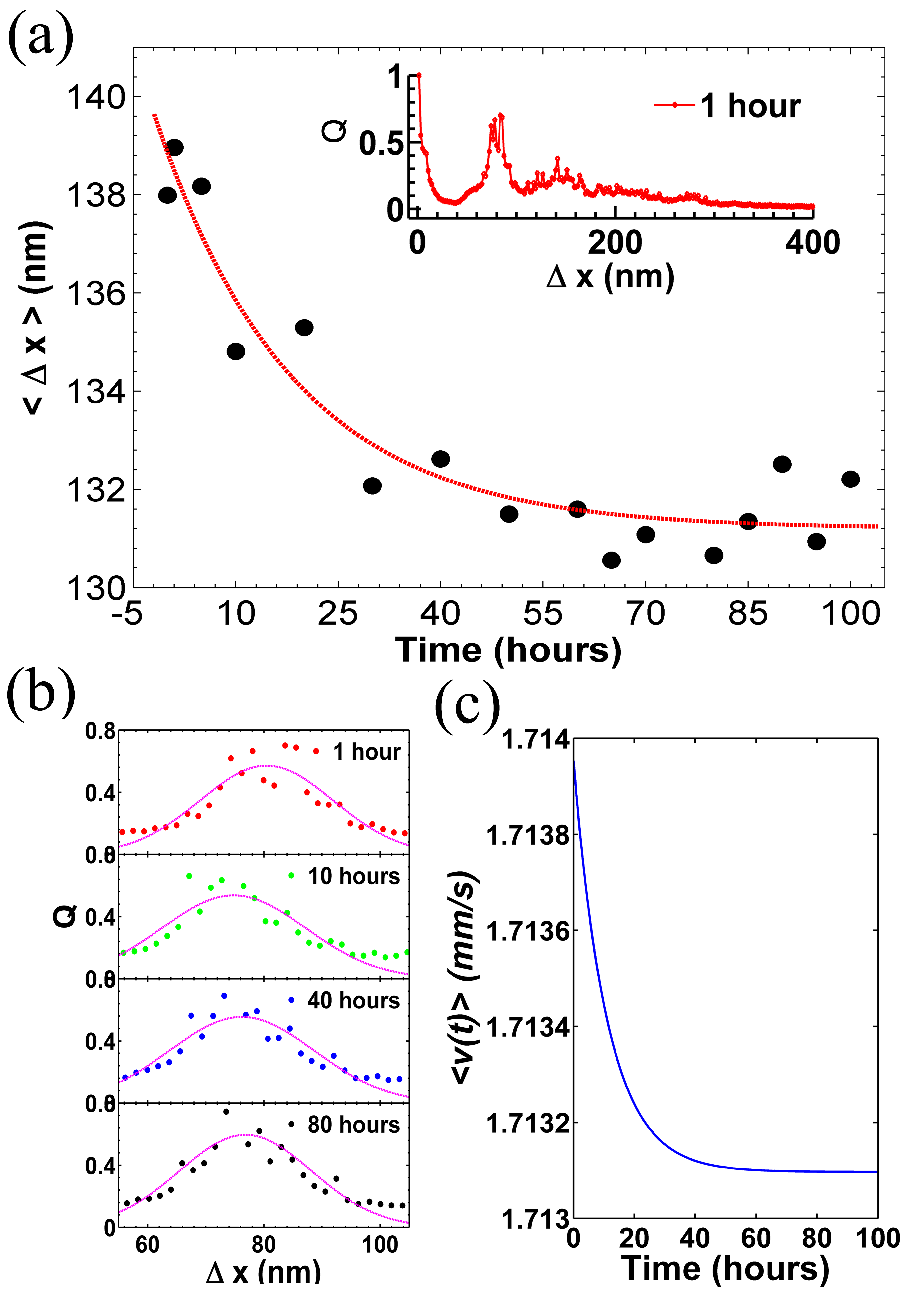}
\caption{The mean displacement $\langle\triangle x\rangle$ decays in the coupling process.
 (a) The average displacement slowly dampens. $\langle\triangle x\rangle$ is calculated from dividing the sum of all displacements (absolute value) in 16.7 ms interval by the total numbers of the steps of particles. The red line is an exponential fit. Inset is the survey spectrum of the displacement at 1 hour. A peak close to zero displacement is mainly caused by the errors of particle tracking. In (a) and (b), \emph{Q} is the normalized quantities.
 (b) Huge quantities of displacements displays a normal distribution centering around 80 nm, showing a statistical maximum probability at a certain value. The normal distribution has been kept in the whole coupling process, which can be fitted with Gaussian function (pinkish lines).
(c) Theoretical predication of the velocity dampening. \emph{k} = 0.9995, $\gamma = 0.000025$.}
\label{Fig2}
\end{figure}

Because the instant velocity of Brownian motion is not available in the present work, here the discussion of the velocity is made on the qualitative assessment. $v_{rms}$ in Fig. 2(c) refers to the instant velocity on a typical time scale, $\sim10^{-6}s$. However, the time scale of the average velocity corresponding to Fig. 2(a) is $\sim10^{-2}$s. The consistency in the dynamic tendency of those two velocities is manifested once the conceptions of the different time scales are reconciled. Consider particle moving $10^{4}$ steps with each step of a time interval $10^{-6}s$, the total duration is $10^{-2}s$. Suppose that the $6\%$ reduction of displacement ($\delta\langle\triangle x\rangle=139nm-131nm$) has been achieved by the accumulation of the reduction over $10^{4}$ steps, with the decrease of each step by $8\times10^{-10}mm$. Thus the velocity reduction is $8\times10^{-10}mm/10^{-6}s$, namely, the velocity dampening from 1 hour to 100 hours is an order of magnitude of $\sim10^{-4}mm/s$, which agrees with the result derived from Eq.(\ref{x1}) shown in Fig. 2(c).

The dynamics of particles reaches the equilibrium state at 100 hours. The ratio of the quantities of $\triangle x$ at various times (Fig. 3(a)) to those at 100 hours is acquired through dividing by the reference spectrum in Fig. 3(b). At the beginning (1 h), large quantities of displacements form the spectrum possessing many spike-like peaks. Thereafter (10 h), these quantities decrease with the whole spectrum lowering towards the base line ($Q_{r}=1$, corresponding value displayed in Fig. 3(b)). With the time proceeding (40 h, 80 h), the quantities of displacements larger than 200 nm continually decrease. Somewhat fluctuation of quantities around the base line appears at 80 hours. Most reduction of the quantities of displacements ($> 80 nm$) occurs in the first 10 hours, whereas small amount of the quantity reduction produces in the subsequent process. In fact, decreasing quantities for displacements ($> 80 nm$) in Fig. 3(a) leads to average displacement decreasing in Fig. 2(a). Both figures indicate that relatively quick dampening occurs in the first 10 hours followed by slow dampening in the subsequent coupling process.

\begin{figure}
\centering
\includegraphics[scale=0.35]{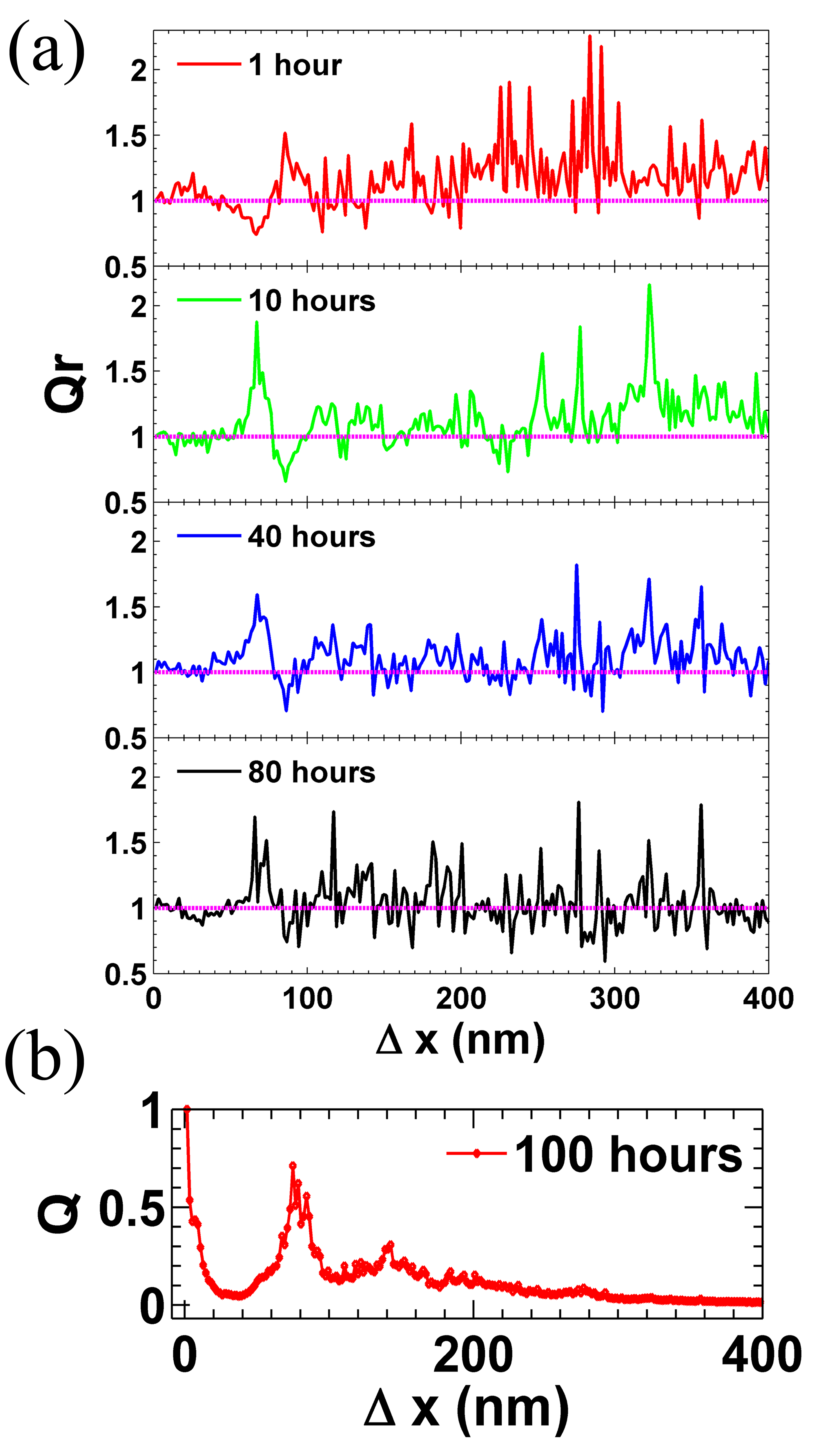}
\caption{Distribution of the particle displacement (in 16.7 ms) at various times.
(a) Decreasing the relative quantities of various $\triangle x$ reflects the dampening of particle motion in the coupling process. The horizontal pinkish line locating at unity value indicates the value in the equilibrium. $Q_{r}$ is the relative quantities, corresponding to the quantities of displacements at different times divided by that at 100 hours in (b).
(b) The spectrum of the displacement in the equilibrium at 100 hours. The peak close to zero displacement is mainly induced by the errors of particle tracking. \emph{Q} is the normalized quantities.}
\label{Fig3}
\end{figure}

The above description indicates the existence of a long quasi-equilibrium of Brownian particles in the qTD confinement, in which the dynamic system slowly proceeds toward the equilibrium. The stochastic force origins from the thermal fluctuation whose effect may be considered invariant. For particles of dampening motion, the friction force and coupling interaction of particles are reducing as velocity decreasing. When the effects of the stochastic force, coupling interaction and friction force cancel out, the confined system is brought to the equilibrium state.

\begin{center}
\textbf{IV. Analysis of coupling motion of particles by statistical mechanics}
\end{center}

Treating the trajectories of particle motion in stochastic process randomly reversing motion direction alone a line, Gaveau \emph{et al.} derived the equation from the statistical mechanics describing the probability density of particle ~\citep{PhysRevLett.53.419}. We find it is feasible to modify that equation by incorporating the damping effect, \emph{i.e.} adding $e^{-\lambda t}$ ($\lambda$, damping coefficient) on both sides of the equation (the derivation see Appendix D),
\begin{equation}\label{a7}
\frac{\partial^{2}P}{\partial t^{2}}+2a_{0}e^{-\lambda t}\frac{\partial P}{\partial t}=(v_{0}e^{-\lambda t})^{2}\frac{\partial^{2}P}{\partial x^{2}}
\end{equation}
where $P(x,t)$ is the probability density of a particle appearing on position \emph{x} at time \emph{t}, $a_{0}$ the initial probability for the change of motion direction in unit time, $v_{0}$ the initial velocity of particle. Solve the equation above, we get (Appendix D),
\begin{equation}\label{a8}
P=v_{0}^{-1}[v_{e}+(v_{0}-v_{e})e^{-\rho t}]e^{-\zeta t}\cos k_{1}x
\end{equation}
in which $k_{1}=\frac{(2n+1)\pi}{2L}$ ($n=0, 1, 2, 3....; \emph{L}$ is the moving distance of the particle without changing direction), $v_{e}$ the velocity at the equilibrium, $\rho=2\sqrt{a_{0}^{2}-k_{1}^{2}v_{0}^{2}}$, $\zeta=(a_{0}-\sqrt{a_{0}^{2}-k_{1}^{2}v_{0}^{2}})e^{-\lambda t}$.

Eq.\ref{a8} indicates that the probability density of particle fluctuates in the spatial distribution. In Brownian motion,  particles move in random directions. The superimposition of many density fluctuations with random directions leads to the fluctuation (inhomogeneity) of particle distribution, as shown in Fig. 1(a). This is analogue to the works of Cahn and Vrij in spinodal decomposition~\citep{cahn1965phase} and spinodal dewetting~\citep{vrij1966possible}, respectively. Cahn and Vrij displayed that the superimposition of fluctuations (of composition/film thickness) with random directions forms the inhomogeneous distribution of concentration~\citep{cahn1965phase} and thickness~\citep{vrij1966possible}. Particle trajectories are produced by the positions of particles accumulated in a time interval, which are still keeping in inhomogeneous distribution (Figs. 4a and 4b).

\begin{figure}
\centering
\includegraphics[scale=0.75]{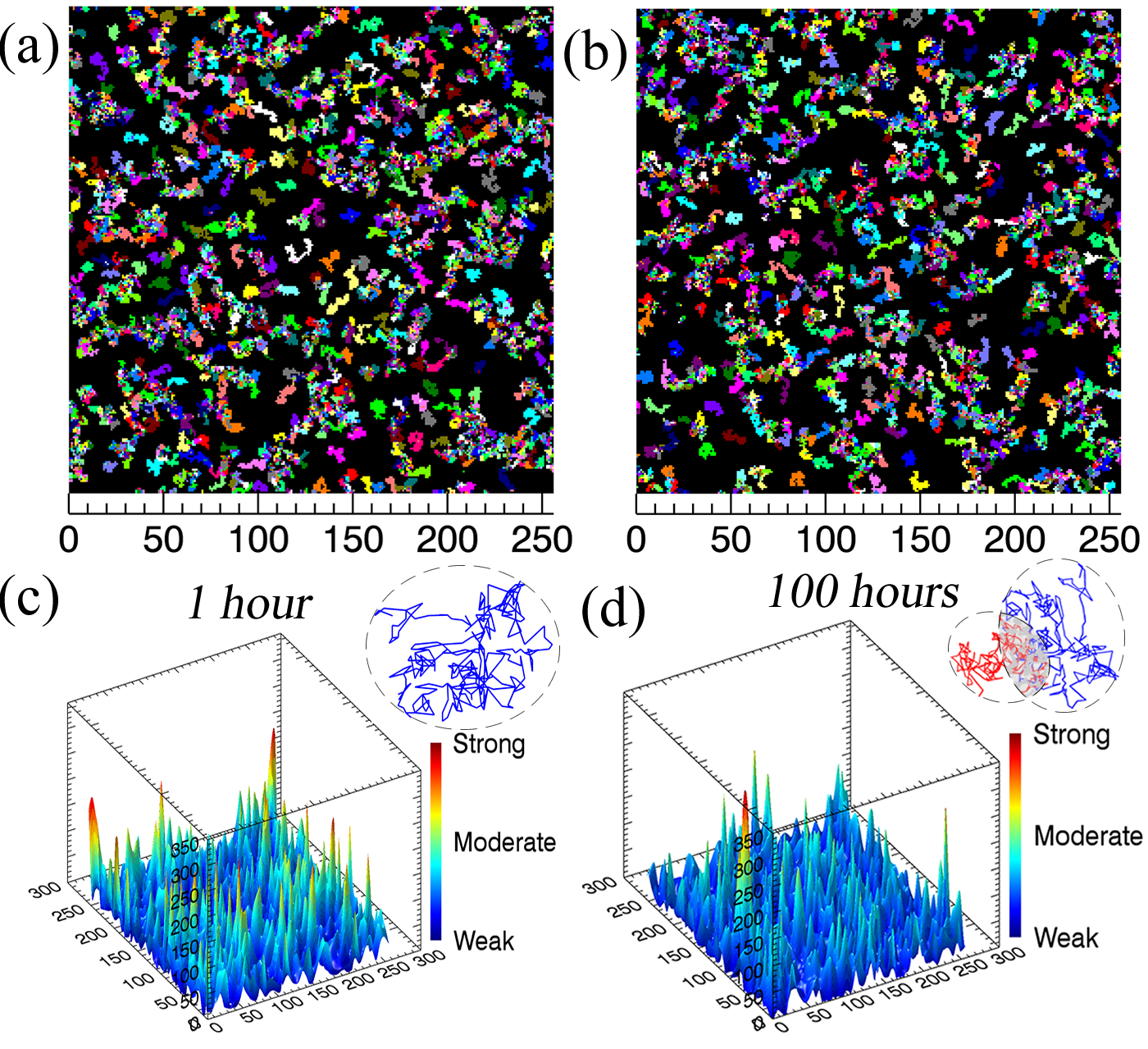}
\caption{Coupling the motions of particles and overlapping the trajectories of motions.
(a) and (b) The entanglement of the trajectories of particles at 1 hour and 100 hours, respectively. Different colors are used to identify the trajectories for different particles. The scale bar shows the view field of $256\times 256$ pixels.
(c) and (d) Overlapping trajectories represented by the peaks. The higher peak corresponds to the strong overlapping of the more trajectories, dictating with warm colors (yellow even red at the peaks). Insets in (c) and (d) are the active area of an individual particle and the illustrative overlapping, respectively.}
\label{Fig4}
\end{figure}

For the particle with the relatively larger velocity, it appears in more locations in the unit time, thus corresponding to the larger probability density here. Eq.\ref{a8} indicates that the probability density of particle decreases with time due to the dampening, corresponding to the reduction of velocity. The dampening factor $\zeta $ exponentially decreases with time, meaning that the dampening effect lessens with the time, which is consistent with the experimental results in Figs. 2a and 3a.

The effect of particle interaction in the coupling process is indicated in the experimental results shown in Fig. 4. Through the analysis of the details (not obvious in Figs. 4a and 4b), it shows that more trajectories of different particles (displayed by different colors) intertwine together in Fig. 4(a) than that in Fig. 4(b), indicating the stronger particle interaction at 1 hour than that at 100 hours. To quantitatively evaluate the interaction, in Figs. 4c and 4d, overlapping trajectories of different particles are represented by the height. The higher peak means that more trajectories are overlapped (\emph{i.e.} stronger intensity of overlapping). The heights of the peaks have reduced after long coupling process, implying that the intensity of trajectories overlapping decreases, \emph{i.e.} particle interaction dampens with time. We denote that the active area of the particle is the area $S$ enclosing the trajectories of individual particle for a time interval $\Delta t$ indicated by the dash circle (Fig. 4(c) inset). The reduction of the heights of peaks results from overlapping active area (Fig. 4(d) inset) decreasing, corresponding to active area shrinking. Thus the probability to find the particle in unit area decreases, \emph{i.e.} the probability density of particle reduces, as corroborated by Eq.\ref{a8}. Therefore, the reduction of overlapped areas means that the intensity of particle interaction has decayed.

\begin{center}
\textbf{V. Conclusion}
\end{center}

In summary, we report the existence of a quasi-equilibrium state of Brownian particles in the qTD confinement. Coupling particle motions induced by the confinement results in slowly dampening the displacement and interaction of particles. The theory developed here predicts the dynamic coupling motions of Brownian particles, which is consistent with the experimental result.\\
\begin{acknowledgments}
\footnotesize
Acknowledgments. We thank Guangda Li for writing the software program for data analysis. The authors gratefully acknowledge the support by NSFC (No.11104218), Natural Science Basic Research Plan in Shaanxi Province of China (Program No.2010JZ001), Scientific Research Foundation for the Returned Overseas Chinese Scholars (Shaanxi Administration of Foreign Expert Affairs, 2011).
\end{acknowledgments}

\begin{center}
\textbf{Appendix A: Experiment}
\end{center}

\textbf{Materials.}
Colloidal suspension of polystyrene microspheres (Microparticles GmbH, Germany) is used as received. Particles were momodispersed in aqueous solution with the weight percent $10 wt\%$. The diameter of particle is $1.39 \mu m$, with the Standard Deviation (SD) of $0.04 \mu m$. The suspension is further diluted into $1 wt\%$ by Deionized (DI) water (1.0 g/mL) with the standard resistivity of 18.29 $\Omega.cm$. To exclude the gravity effect of particles, heavy water ($D_{2}O$: density, 1.107 g/mL; purity, $99.9\%$; Sigma-Aldrich) and DI water is mixed to make the aqueous solution with the density of 1.05 g/mL, the same as the density of the polystyrene microspheres.

\textbf{Experimental setup.}
Two parallel fused quartz plates (UQG Optics, UK) with the dimension of $25mm \times 25mm \times 1mm$ are horizontally placed as a qTD cell. The quartz plates are polished into the high level uniformity of the roughness and flatness by the product supplier. Root mean square roughness of both sides of the plates is characterized by Atomic Force Microscopy (Dimension Icon, Veeco) as ~0.6 nm in $10 \mu m \times 10 \mu m$ area. Flatness and parallelism are better than 5 optical fringes and less than 3 minute arc (1/20 degree), respectively. The quartz plates are washed subsequently by acetone, ethanol and DI water, then dried by dry nitrogen flow.

 Thanks to hydrophobic property of the plates, $0.4 \mu L$ polystyrene suspension is dropped on the bottom substrate. Silicone oil is coated on the outer area of the droplet to form a 'corral'. Silica microspheres (diameter: 7.75 $\mu m$, SD = 0.29 $\mu m$, $5 wt\%$ aqueous suspension, GmbH, Germany) are used as the spacers on the plate edges to separate the two plates.

To prevent the evaporation and the leaking of water throughout the observation, optical adhesive (NOA 63, Norland, USA) is useful to seal the cell. The adhesive is exposed by 365 nm ultraviolet light with 3W power for 3 minutes. The cell then cools down for about 1 minute.

\textbf{Optical observation.} Long working distance optical microcopy (Eclipse LV100, Nikon) with the transition mode and objective lens (50x, CFI, ELWD, NA=0.55, WD=9.8 mm) is employed to image the particle motion. Marlin AVT CCD camera is operating with 60 frames/s and the resolution of 475 nm/pixel. The depth of field for the optical system is about $1 \mu m$, which is smaller than the diameter of the particle devoting to the good focus on one layer of the particle in the solution. Due to mass storage problem, particle motion has be recorded for ten minutes at the beginning of every hour, then the illumination is turned off to avoid thermal perturbation on the cell. The temperature during the observation is relative stable (with the variation less than $2\textordmasculine C$).

\textbf{Image processing.} A plugin in ImageJ from NIH developed by the MOSAIC Group in Max Planck Institute of Molecular Cell Biology and Genetics is employed to detect the particles and track their motion in the 2D surface. Specific parameters suitable for our system is: image imported: 1000 frames; view field: $256\times256$ pixels; time interval: $\tau=16.7 ms$; radius of the particle:~ 3 pixels; cut-off: 0; percentile: 2; link-frame: 2 frames; displacement: 1 pixel. By running the plugin, the trajectory for each particle is recognized and recorded as coordinates \emph{x}-\emph{y}. The adjacent coordinates are used to calculate the mean displacement $\langle\triangle x\rangle$ by summing the individual displacement $\Delta X_{i,j}$ and making the average of all displacements,
\begin{equation}\label{n1}
\Delta X_{i,j}=((x_{i,j}-x_{i,j-1})^{2}+(y_{i,j}-y_{i,j-1})^{2})^{0.5}
\end{equation}
where \emph{i} is trajectory sequence, and \emph{j} is position sequence of one particle for its trajectory. Trajectories overlapping is calculated as,
\begin{equation}\label{n2}
C_{x}^{i}=\frac{1}{m}\sum _{j=1}^{m}x_{i,j},  C_{y}^{i}=\frac{1}{m}\sum _{j=1}^{m}y_{i,j}
\end{equation}
where $C_{x}^{i}$ and $C_{y}^{i}$ are the center coordinates of the $i^{th}$ trajectory, \emph{m} is the number of the positions in a time duration. For example, when the particle \emph{A} and \emph{B} are very close, their trajectories will overlap as time going, then the position of the center of the trajectory \emph{A} is apart from that of the trajectory \emph{B} within a critical distance $d_{0}$. We define
\begin{equation}\label{n3}
d=((C_{x}^{i}-C_{x}^{i-1})^{2}+(C_{y}^{i}-C_{y}^{i-1})^{2})^{0.5}
\end{equation}
as the distance between the centroids of two trajectories. Parameter $z$ is introduced to measure the degree of coupling strength of two trajectories, namely the interaction between two particles, so that,
\begin{equation}\label{n4}
z=0, d\geq d_{0}; z=k(d_{0}-d), d<d_{0}
\end{equation}
where $k$ is the interaction factor and set to 10 in present case, and $d_{0}$ is the diameter of particle which corresponds to the minimum distance of the centers of two adjacent particles.

\begin{center}
\textbf{Appendix B: Collision Rate}
\end{center}
\  The collision rate can be qualitatively analyzed by approximately considering the bulk and qTD suspension consisted of the cube and square zones (with colloidal particles as vertices, Fig. 1(c) inset), respectively. The number of particle in a cube: $\frac{1}{8}\times 8 = 1$; thus the volume density of particle $\varrho_{b}=\frac{4\pi r^{3}}{3a^{3}}$, where $a$ is lattice length. After a particle has passed a cube zone, the volume of its path $V_{b} \approx \pi r^{2}a$. So the possibility of the particle colliding with the other particles (at vertices) in a cube, $p_{b}=\frac{4\pi^{2}r^{5}}{3a^{5}}$.
The number of particle in a 2D square area: $\frac{1}{4}\times 4=1$; so the area density of particle $\varrho_{q}= \frac{\pi r^{2}}{a^{2}}$. The area of the path corresponding to particle passing through the square zone, $S_{q} \approx 2ra$. Then the possibility of the particle colliding with other particles (at vertices) in a square, $p_{q} = \frac{2\pi r^{3}}{a^{3}}$. Thus $p_{q}/p_{b} = 0.5(\frac{a}{r})^{2}\gg 1$. Therefore, as the particle passes the same distance in the suspension, the possibility of particle collision in the qTD confinement is much large than that of the bulk.

\begin{center}
\textbf{Appendix C: Solution of Eq.\ref{a4}}
\end{center}
Solve Eq.\ref{a4}, we get,
\begin{equation}\label{w1}
v=(k-1+e^{\gamma t})^{-\frac{\beta}{\gamma}}\{{\int^{t}_{0} [1+(k-1)e^{-\gamma t}]^{\frac{\beta}{\gamma}-1}e^{\beta t}A(t)dt+C}\}
\end{equation}
  $\gamma$ is far less than and \emph{k} is very close to unity, respectively, thus, $(k-1)e^{-\gamma t}\ll 1$. With series expansion, $\int^{t}_{0}[1+(k-1)e^{-\gamma t}]^{\frac{\beta}{\gamma}-1}e^{\beta t}A(t)dt\approx \int^{t}_{0} [1+(\frac{\beta}{\gamma}-1)(k-1)e^{-\gamma t}]e^{\beta t}A(t)dt=\int^{t}_{0} [e^{\beta t}+(\frac{\beta}{\gamma}-1)(k-1)e^{(\beta-\gamma)t}]A(t)dt$, and $\beta\gg \gamma$, thus $[1+(k-1)e^{-\gamma t}]^{\frac{\beta}{\gamma}-1}$ can be moved outside the integral symbol with negligible error on the integral value,
\begin{equation}\label{w2}
\begin{array}{r@{~}l}
&\int^{t}_{0}[1+(k-1)e^{-\gamma t}]^{\frac{\beta}{\gamma}-1}e^{\beta t}A(t)dt\approx\\
&[1+(k-1)e^{-\gamma t}]^{\frac{\beta}{\gamma}-1}\int^{t}_{0} e^{\beta t}A(t)dt
\end{array}
\end{equation}
For initial condition $t=0$ and $v(0)= v_{0}$, from Eq.\ref{w1}, it gets  $C= k^{\frac{\beta}{\gamma}}v_{0}$. Substitute \emph{C} and Eq.\ref {w2} into Eq.\ref{w1}, we get,
\begin{equation}\label{M1}
v=v_{0}k^{\frac{\beta}{\gamma}}(e^{\gamma t}+k-1)^{-\frac{\beta}{\gamma}}+\frac{e^{-\beta t}}{1+(k-1)e^{-\gamma t}}\int^{t}_{0}e^{\beta \xi}A(\xi)d\xi
\end{equation}
Take the square of Eq.\ref{M1} and do the calculation, we get the mean-square velocity $\langle v^{2}\rangle$ as presented in Eq.\ref{a5}.\\
Integrate Eq.\ref{M1} to get the displacement,
\begin{equation}\label{M3}
\begin{array}{r@{~}l}
\Delta x =&v_{0}k^{\frac{\beta}{\gamma}}\int^{t}_{0}(e^{\gamma t}+k-1)^{-\frac{\beta}{\gamma}}dt+\\
&\int^{t}_{0}\frac{e^{-\beta t}}{1+(k-1)e^{-\gamma t}}dt\int^{t}_{0}e^{\beta \xi}A(\xi)d\xi
\end{array}
\end{equation}
Due to slow damping process, $\gamma t < 1$,
\begin{equation}\label{M4}
e^{\gamma t}= 1+\gamma t +\frac{1}{2}(\gamma t)^{2}+...\approx1+\gamma t
\end{equation}
With $(k-1)e^{-\gamma t}\ll 1$, $\int^{t}_{0}\frac{e^{-\beta t}}{1+(k-1)e^{-\gamma t}}dt\approx \int^{t}_{0}[{1-(k-1)e^{-\gamma t}}]e^{-\beta t}dt=\int^{t}_{0}[e^{-\beta t}-(k-1)e^{-(\beta+\gamma)t}]dt$, and $\beta\gg \gamma$, thus $\frac{1}{1+(k-1)e^{-\gamma t}}$ can be moved outside the integral symbol with negligible error on the integral value,
\begin{equation}\label{w3}
\int^{t}_{0}\frac{e^{-\beta t}}{1+(k-1)e^{-\gamma t}}dt\approx \frac{1}{1+(k-1)e^{-\gamma t}}\int^{t}_{0}e^{-\beta t}dt
\end{equation}
Substitute Eq.\ref{M4} and Eq.\ref{w3} into the first and second terms of Eq.\ref{M3}, respectively. Integrate the first term and perform integration by parts for the second term, it reaches,
\begin{equation}\label{M5}
\begin{array}{r@{~}l}
\Delta x =&\frac{k v_{0}}{\beta-\gamma}[1-(\frac{k}{\gamma t+k})^{\frac{\beta}{\gamma}-1}]+\frac{1}{\beta[1+(k-1)e^{-\gamma t}]}\\
&[-e^{-\beta t}\int^{t}_{0}e^{\beta \xi}A(\xi)d\xi+\int^{t}_{0}A(\xi)d\xi]
\end{array}
\end{equation}
Square Eq.\ref{M5} and perform the calculation ($\beta\gg\gamma$), we have Eq.\ref{a6}.

\begin{center}
\textbf{Appendix D: Solution of Eq.\ref{a7}}
\end{center}

(a) Derivation of the coefficients of Eq.\ref{a7}. The velocity of particle slowly dampens as $v(t)=v_{0}e^{-\lambda t}$. Consider that a particle has been persistently bombarding by liquid molecules, with the average number of collisions $n_{c}$ per unit length of its trajectory. The particle changes the direction of motion after passing a average length of trajectory \emph{L}, \emph{i.e.} after $n_{c}\emph{L}$ collisions with liquid molecules. The probability for the change of motion direction in unit time, $a=v(t)/L=v_{0}e^{-\lambda t}/L=a_{0}e^{-\lambda t}.$

(b) Set $P(x, t) = X(x)T(t)$ and use the method of separating variables, we convert Eq.\ref{a7} into the following equations,
\begin{equation}\label{M9}
X''+k_{1}^{2}X=0
\end{equation}
\begin{equation}\label{M10}
T''+2a_{0}e^{-\lambda t}T'+k_{1}^{2}v_{0}^{2}e^{-2\lambda t}T=0
\end{equation}
The general solution of Eq.\ref{M9} is $X=c_{1}\cos k_{1}x+c_{2}\sin k_{1}x$. Use the initial condition $P(0, 0) = X(0)T(0)=1$, as $T(0)=1$ (no damping at $t=0$), $X(0)=1$, get $X(0)=c_{1}=1$. Furthermore, $\frac{dX}{dx}|_{x=0}=0$, it has $c_{2}k_{1}=0, c_{2}=0$. In addition, $P(L,0)=X(L)T(0)=0, X(L)=0$, so $X(L)=\cos k_{1}L=0, k_{1}=\frac{(2n+1)\pi}{2L}, n=0,1,2...$. Thus we have,
\begin{equation}\label{M11}
X=\cos k_{1}x
\end{equation}
We realize that $T_{1}=e^{-st}$ can be a particular solution of Eq.\ref{M10}. Substitute $T_{1}$ into the equation and get, $s_{1,2}=e^{-\lambda t}(a_{0}\pm\sqrt{a_{0}^{2}-k_{1}^{2}v_{0}^{2}})$. In case of the dampening, we take $s=e^{-\lambda t}(a_{0}-\sqrt{a_{0}^{2}-k_{1}^{2}v_{0}^{2}})=c_{0}e^{-\lambda t}$. The second linearly independent solution of Eq.\ref{M10} is, $T_{2}=T_{1}\int(e^{-\int2a_{0}e^{-\lambda t}dt}/T_{1}^{2})dt=T_{1}\int e^{2e^{-\lambda t}(\frac{a_{0}}{\lambda}+c_{0}t)}dt$. As $\lambda$ has the similar character as $\gamma$ in Eq.\ref{M4}, thus, $T_{2}\approx T_{1}\int e^{2(1-\lambda t)(\frac{a_{0}}{\lambda}+c_{0}t)}dt=\frac{T_{1}}{2(c_{0}-a_{0})}e^{2[\frac{a_{0}}{\lambda}+(c_{0}-a_{0})t]}$. So we get,
\begin{equation}\label{M12}
T(t)=c_{3}T_{1}+c_{4}T_{2}=e^{-c_{0}te^{-\lambda t}}(c_{3}+c_{5}e^{-\rho t})
\end{equation}
where $c_{5}=\frac{c_{4}}{2(c_{0}-a_{0})}e^{\frac{2a_{0}}{\lambda}}, \rho=2\sqrt{a_{0}^{2}-k_{1}^{2}v_{0}^{2}}, c_{0}=a_{0}-\sqrt{a_{0}^{2}-k_{1}^{2}v_{0}^{2}}$.
The probability density in an interval $[x, x+dx]$ on the line of length \emph{L} is proportional to the numbers counted for a particle lying in $[x, x+dx]$. The larger the velocity is for the particle, the more counts it has in unit time, \emph{i.e.} $P\propto v$. The ratio of probability density for $t\rightarrow\infty$ to that for $t=0$, $\frac{P(x, \infty)}{P(x, 0)}=\frac{X(x)T(\infty)}{X(x)T(0)}=\frac{T(\infty)}{T(0)}=\frac{v_{e}}{v_{0}}$, where $v_{e}$ is the velocity at the final equilibrium state as $t\rightarrow\infty$. Since$\underset{t\rightarrow\infty}{\lim}-c_{0}te^{-\lambda t}=\underset{t\rightarrow\infty}{\lim}\frac{-c_{0}t}{e^{\lambda t}}=0$, from Eq.\ref{M12}, $T(\infty)=c_{3}= v_{e}/v_{0}$. For the initial condition $T(0)=c_{3}+c_{5}=1$, $c_{5}=1-\frac{v_{e}}{v_{0}}$. We get,
\begin{equation}\label{M13}
T(t)=v_{0}^{-1}e^{-c_{0}te^{-\lambda t}}[v_{e}+(v_{0}-v_{e})e^{-\rho t}]
\end{equation}
From Eqs.\ref{M11},\ref{M13}, we thus arrive at the product solution Eq.\ref{a8}.


\begin{thebibliography}{25}

\begin{center}
\textbf{References}
\end{center}

\bibitem{huang2012nano}
M.-J. Huang, H.-Y. Chen, and A. S. Mikhailov, Eur. Phys. J. E \textbf{35,} 119 (2012).

\bibitem{tan2012hydrodynamic}
M. Tan, D. Le, and K. Chiam, Soft Matter \textbf{8,} 2243 (2012).

\bibitem{bonilla2012hydrodynamic}
B. Bonilla-Capilla, A. Ram\'{\i}rez-Saito, M. Ojeda-L\'{o}pez, and J. Arauz-Lara, J. Phys.: Condens. Matter. \textbf{24,} 464126 (2012).

\bibitem{novikov2010hydrodynamic}
S. Novikov, S. A. Rice, B. Cui, H. Diamant, and B. Lin, Phys. Rev. E \textbf{82,} 031403 (2010).

\bibitem{uspal2012scattering}
W. E. Uspal and P. S. Doyle, Phys. Rev. E \textbf{85,} 016325 (2012).

\bibitem{C2SM25931A}
W. E. Uspal and P. S. Doyle, Soft Matter \textbf{8,} 10676 (2012).

\bibitem{nugent2007colloidal}
C. R. Nugent, K. V. Edmond, H. N. Patel, and E. R. Weeks, Phys. Rev. Lett. \textbf{99,} 025702 (2007).

\bibitem{edmond2012influence}
K. V. Edmond, C. R. Nugent, and E. R. Weeks, Phys. Rev. E \textbf{85,} 041401 (2012).

\bibitem{eral2009influence}
H. B. Eral, D. van den Ende, F. Mugele, and M. H. G. Duits, Phys. Rev. E \textbf{80,} 061403 (2009).

\bibitem{carbajal1997brownian}
M. D. Carbajal-Tinoco, G. Cruz de Le\'{o}n, and J. L. Arauz-Lara, Phys. Rev. E \textbf{56,} 6962 (1997).

\bibitem{cui2004anomalous}
B. Cui, H. Diamant, B. Lin, and S. A. Rice, Phys. Rev. Lett. \textbf{92,} 258301 (2004).

\bibitem{diamant2009hydrodynamic}
H. Diamant, J. Phys. Soc. Jpn. \textbf{78,} 1002 (2009).

\bibitem{crocker1996methods}
J. C. Crocker and D. G. Grier, J. Colloid Interface Sci., \textbf{179,} 298 (1996).

\bibitem{uhlenbeck1930theory}
G. E. Uhlenbeck and L. S. Ornstein, Phys. Rev., \textbf{36,} 823 (1930).

\bibitem{peyla2011new}
P. Peyla and C. Verdier, EPL \textbf{94,} 44001 (2011).

\bibitem{batchelor1972hydrodynamic}
G. Batchelor and J. Green, J. Fluid Mech. \textbf{56,} 375-400 (1972).

\bibitem{einstein1956investigations}
A. Einstein, \emph{Investigations on the Theory of the Brownian Movement} (R. F\"{u}rth, Ed. Dover Publications, New York, 1956).

\bibitem{PhysRevLett.53.419}
B. Gaveau, T. Jacobson, M. Kac, and L. S. Schulman, Phys. Rev. Lett. \textbf{53,} 419 (1984).

\bibitem{cahn1965phase}
J. W. Cahn, J. Chem. Phys. \textbf{42,} 93 (1965).

\bibitem{vrij1966possible}
A. Vrij, Discuss. Faraday Soc. \textbf{42,} 23 (1966).

\end{thebibliography}
\end{document}